\documentclass[12pt]{article}
\usepackage{amssymb}

\newcommand{\cA}{{\cal A}}

\newcommand{\cC}{{\cal C}}

\newcommand{\cG}{{\cal G}}
\newcommand{\cH}{{\cal H}}

\newcommand{\cS}{{\cal S}}

\newcommand{\bo}{{\bar o}}

\newcommand{\bM}{{\bar M}}

\def\a{\alpha}
\def\b{\beta}

\def\s{\sigma}

\def\ie{{\it i.e.}}

\def\ZZ{\mZ}
\def\bo{{\raise.15ex\hbox{\large$\Box$}}}               

\def\face{{\raise.2ex\hbox{$\displaystyle \bigodot$}\mskip-2.2mu \llap {$\ddot
        \smile$}}}                                      

\def\leftrightarrowfill{$\mathsurround=0pt \mathord\leftarrow \mkern-6mu
        \cleaders\hbox{$\mkern-2mu \mathord- \mkern-2mu$}\hfill
        \mkern-6mu \mathord\rightarrow$}       
\def\dvec#1{\vbox{\ialign{##\crcr
        \leftrightarrowfill\crcr\noalign{\kern-1pt\nointerlineskip}
        $\hfil\displaystyle{#1}\hfil$\crcr}}}           

\def\beq{\begin{equation}}
\def\eeq{\end{equation}}

\def\beqx{\begin{displaymath}}
\def\eeqx{\end{displaymath}}

\def\beql{\begin{eqnarray}}
\def\eeql{\end{eqnarray}}

\newcommand{\Tr}{{\rm Tr}}

\newcommand{\bea}{\begin{eqnarray}}
\newcommand{\eea}{\end{eqnarray}}

\newcommand{\R}[1]{(\ref{eq:#1})}
\newcommand{\mod}{\;\;\;\;{\rm mod }\;}

\newcommand{\ii}{{\rm i}}

\def\[{\left [}
\def\]{\right ]}
\def\({\left (}
\def\){\right )}

\def\ZZ{\mathbb{Z}}

\def\cA{{\cal A}}  \def\cC{{\cal C}}
  
\def\cG{{\cal G}} \def\cH{{\cal H}}

\def\cS{{\cal S}}

\def\+{\oplus}

\begin{document}

\vspace*{0.15in}
\hbox{\hskip 12cm NIKHEF/2004-002  \hfil}
\hbox{\hskip 12cm KUL-TF-04/07  \hfil}
\hbox{\hskip 12cm hep-th/0403196 \hfil}
\begin{center}
{\Large \bf Chiral Supersymmetric Standard Model Spectra\\}

\vspace*{0.10in}

{\Large \bf from Orientifolds of Gepner Models}

\vspace*{0.5in}
{T.P.T. Dijkstra${\,}^\Diamond$\footnote{email:tdykstra@nikhef.nl},
L. R. Huiszoon${\,}^\dagger$\footnote{email:lennaert@itf.fys.leuven.ac.be} 
\& A.N. Schellekens${\,}^\Diamond$\footnote{email:t58@nikhef.nl}}\\[.3in]
{\em
$\Diamond \;\; $ NIKHEF Theory Group \\
	Kruislaan 409 \\
	1098 SJ Amsterdam \\
	The Netherlands \\
$\dagger \;\; $ Instituut voor Theoretische Fysica\\
	Katholieke Universiteit Leuven\\
	 Celestijnenlaan 200D\\
	B-3001 Heverlee \\
	Belgium}\\[0.2in]

\end{center} 

\begin{center}
\vspace*{0.20in}
{\bf Abstract}

\end{center}

We construct $d=4,N=1$ orientifolds of Gepner models with just the chiral spectrum of the
standard model. We consider all simple current modular invariants of
$c=9$ tensor products of $N=2$ minimal models. For some very specific tensor
combinations, and very specific modular invariants and orientifold projections,
we find a large number of such spectra.\footnote{Updated and almost complete results
are available at www.nikhef.nl/$\sim$t58/Orientifold\_results.ps.} We allow for standard model singlet (dark) matter and
non-chiral exotics. The Chan-Paton
gauge group is either
$U(3) \times Sp(2) \times U(1) \times U(1)$ or $U(3) \times U(2) \times U(1) \times U(1)$.
In many cases the standard model hypercharge $U(1)$ has no coupling to RR 2-forms and hence
remains massless; in some of those models the $B\!-\!L$ gauge boson does acquire a mass.

\newpage

\section{Introduction}   

There are many possible ways in which the Standard Model might emerge from String Theory.
One of them is a standard gauge unification scenario using the Heterotic string as a starting point.
Another broad class, with particular advantages described extensively in many
 papers~\cite{BlGKL0007,AlFIRU0011,IbMR0105,BlKLO0107,CvSU0107,BaKL0108,BlBKL0206,Ko0207,BlGO0211,
CrIM0302,H0303,CaU0303,LaP0305,CvLL0403}\footnote{For reviews consult~\cite{Ur0301,Ot0309,K0310,Lu0401}.}, is through
intersecting stacks of branes. In both cases an important issue is to find various kinds
of (semi)-realistic examples. Here we will focus on a class of examples that was rather difficult
to obtain so far, namely supersymmetric spectra that satisfy all the tadpole cancellation conditions. 
There are some results~\cite{CvSU0107,CvPS0212,CvP0303} in this area using orientifolds of toric orbifolds~\cite{BiSa,GiP}, 
but we want to consider here internal CFTs that are non-trivial~\cite{Cardy,Sa}.

We consider all $d=4,N=1$ simple current orientifolds of Gepner models 
(see~\cite{BlW9806,GoM0306,AlALN0307,Bl0310,BrHHW0401, BlW0401} for some specific cases including chiral spectra.) 
At this point in moduli space, the underlying
$d=2$ conformal field theory is a $c=9$ tensor combination of $N=2$ minimal models.
These CFTs are rational, so the results of~\cite{FOE} apply to construct and classify 
$\a'$-exact orientifold vacua.

In this set of M-theory vacua, we are performing a
systematic search for finite, perturbatively stable points with just the chiral spectrum of the standard model.
We allow for non-chiral matter in standard model gauge group representations, 
and for any matter in hidden gauge group representations. 
The total Chan-Paton
 group includes $U(3)_a\times U(2)_b \times U(1)_c\times U(1)_d$ 
or $U(3)_a\times Sp(2)_b \times U(1)_c\times U(1)_d$. The standard model particles are basically realized as 
 in~\cite{IbMR0105}. The number of Higgses is left free.
  Such models are notoriously hard to construct in orientifolds of (orbifolded) tori. 
In this paper we show that this is not the case in the class of models under
consideration here, provided one chooses just the right modular invariant partition function and orientifold
projection. 
 
In section~\ref{sec-SCGM} we review the material presented in~\cite{FOE}
and formulate our precise search criteria. In section~\ref{sec-models} we report our results.
For some tensor combination
of minimal models, like $(6,6,6,6)$, we have found thousands of inequivalent spectra satisfying our conditions.
In order to avoid endless tables, we will briefly discuss
 a few spectra that are chosen from this set.

\section{Orientifolds of Simple Current Gepner Models} \label{sec-SCGM}

A fundamental property of D-branes and O-planes is that they are
defined in terms of boundary conditions for strings~\cite{Po}. This means that these objects can be studied by
CFT methods, in which the relevant objects are called boundary and crosscap states. These encode the brane/plane
tension and RR charges, as well as the perturbative spectrum of string vibrations in a orientifold vacuum. 

Based on earlier work of~\cite{SaS,P,SBB1,SBB2,klein} a general formula was presented in~\cite{FOE}
 for the boundary and crosscap states
for a large class of rational CFTs. 
This class is non-trivial in any RCFT that has simple currents. Such currents are present in 
abundance in many cases of interest, and in particular in (tensor products of) $N=2$
minimal models. In \cite{GaS92,KrS} a classification of such invariants was obtained. In \cite{FOE}
a description was given of the simple current based orientifolds of these invariants.  Here we
consider all these simple current invariants and all orientifolds described, applied
to the $c=9$ tensor products of $N=2$ minimal models \cite{Gepner:1988qi}.

The ``internal" CFT built out of $N=2$ minimal models is tensored with the space-time NSR sector.
We view all factors in the tensor product as non-supersymmetric CFTs. In order to obtain a CFT
with global world sheet supersymmetry, the chiral algebra must be extended; a second
extension is needed to obtain $N=1$ supersymmetry in target space. In order to describe these
extensions it is convenient to replace the NSR sector by a bosonic CFT, namely a $SO(10)$ 
level one affine Lie algebra, using the
``bosonic string map" (see ~\cite{LeSW} and references therein, and ~\cite{FSW,HS} for 
recent applications of this method).
Hence we consider
\beq
\cA_{tensor} = D_{5,1} \otimes_{i=1}^r \cA_{k_i}
\eeq
where $\cA_k$ is the $N=2$ minimal model at level $k$ with conformal anomaly
\beq
c_k = \frac{3k}{k+2} \;\;\; .
\eeq
The constraint $\sum_i^r c_{k_i} = 9$ leads to $168$ inequivalent tensor products $\cA_{tensor}$.
In each factor we denote the supercurrent as ``$v$" (in the NSR sector this is actually a vector that
acquires conformal weight $\frac32$ by multiplication with $\partial X^{\mu}$); in each factor this is
a simple current. Furthermore each factor contains two Ramond simple currents, which we denote as ``$s$" and ``$c$".
In order for $\cA_{tensor}$ to be $N=2$ world-sheet supersymmetric, we need to extend the algebra by the
fermion allignment simple currents $(v,v,0,0,...),(v,0,v,0,...), ... $. The resulting algebra is called
$\cA_{ws}$. As a result of this extension, all primaries of $\cA_{ws}$ are either in the
Ramond (R) or Neveu-Schwarz (NS) sector. Space-time supersymmetry is obtained when we
extend by the spectral flow simple current $(s,s,s,s,...)$ that relates the R and NS sector. This extension is
called $\cA_{st}$ and is our starting point. Typically it has several thousands primaries, a few
tens or hundreds of which are simple currents. We consider all symmetric simple current invariants
generated by the formula of \cite{KrS}. These invariants may be of automorphism type, or of extension
type, or any combination thereof. The heterotic string spectra of these CFTs 
and their modular invariant partition functions (MIPFs) have
been intensively scanned in the past \cite{Lutken:1988zj,Lynker:fs,Schellekens:1989wx,
Fuchs:1989pt,Fuchs:1990yv,Greene:1990ud,Kreuzer:1993uy}
resulting in a large number of spectra with three chiral families
only for the combination (1,16,16,16) \cite{Gepner:1987hi} and some of its modular invariants \cite{Schellekens:1989wx}.

In all cases we consider the complete set of boundaries. In the
case of extension modular invariants this includes boundaries that do not respect the extended symmetries,
as required to fulfill the completeness condition of \cite{Pradisi:1996yd}. Note however, that all boundaries and crosscaps
respect the symmetries of $\cA_{st}$, and in particular the same copy of $N=1$ target space supersymmetry. The formalism
could equally well be applied to subalgebras of $\cA_{st}$, such as  $\cA_{ws}$, in which case we would
be able to consider models with brane supersymmetry breaking \cite{Antoniadis:1999xk}. 
 However, due to the huge number
of primaries and boundaries this is computionally more challenging, and will not be considered here.

An important ingredient in these computations is the resolution of simple current fixed points
that occur for all even values of $k$. This enters the computation at two points, namely for
obtaining the modular $S$-matrix of $\cA_{st}$, and in the computation of the boundary coefficients
for non-trivial MIPFs. A general formula for $S$ for simple current extended tensor products and
coset CFTs was derived in \cite{FSS}. In addition to this we need a formula for the simple current
fixed point resolution matrices $S^J$ for all simple currents $J$ of $\cA_{st}$. This formula was
derived in \cite{Schellekens:1999yg}. In the case we  consider here, all these fixed point resolution matrices
are of course related to those of $SU(2)$ level $k$, which is just a number, but there are several
non-trivial phases to keep track of, that originate from field identification in the minimal models,
and the extensions that lead to $\cA_{st}$. Once these matrices are available, all cases are 
equally easy to deal with as the ``Cardy case" (the charge conjugation modular invariant).

We start by reviewing the construction of Simple Current Gepner Models. Then we present a canonical class of
boundary and crosscap states for these theories and write down the tadpole conditions. 
The spectrum can then be calculated as reviewed in~\cite{SaS,AnS,thesis}. At the end of this section we formulate
our search criteria.

The following notation is understood. Chiral primaries are denoted by $i,j$
and there characters by $\chi_i$ and conformal weight by $h_i$. 
The superscript in $i^c$ denotes $d=2$ charge conjugation. 
 Simple currents are denoted by $J,K,L$ and their order by $N_J,...$. 
 The monodromy charge of $i$ with respect to $J$ is $Q_J(i)$. 
Simple current groups are
denoted like $\cH$ and the number of elements by $|\cH|$. The rest of our
notation will be explained in the text.

\subsection{Simple Current Gepner Models}

A single $N=2$ minimal model has simple current group $G_k = \ZZ_{4k}$ when $k$ odd and $\ZZ_{2k}\times \ZZ_2$
when $k$ even.
The extended algebra $\cA_{st}$ has a remaining simple current group $\cG_{st}$, whose structure depends on the
details of the model. For every subgroup~\footnote{In addition, all elements of $\cH$ must satisfy the condition
that spin times order is integral.} $\cH\in \cG_{st}$, and a matrix $X$, defined modulo integers, that obeys
\bea
X(J,K) + X(K,J) & = & Q_J(K) \mod 1 , J\neq K \\
 X(J,J) & =&  -h_{J} \mod 1
\eea
plus the constraints $N_J X(J,K)\in \ZZ$, $X(J,K) N_K \in \ZZ$,
we can define string vacua with modular invariant torus partition function
\beq \label{eq:torus}
Z(\cH,X) = \sum_{i,j} \chi_i \chi_{j^c} Z_{ij} \;\;\; ,
\eeq
where $Z_{ij}$ is the number of currents $J\in \cH$ such that
\bea
j & = & Ji \\
 Q_K(i) + X(K,J) & = & 0 \mod 1 \;\;\; 
\eea
for all $K \in \cH$.
In this language, the ordinary ``Gepner model" corresponds to the choice $\cH=\{0\}, X=0$, i.e.
 the charge conjugation invariant of $\cA_{st}$. The number of invariants obtained in this way
 grows rapidly with the number of cyclic factors in $\cG_{st}$.

\subsection{Boundary and Crosscap States}

We now present the results of ~\cite{FOE} in a slightly modified form that is more suitable
for our purposes. This only involves some reshuffling of phase factors in the coefficients; the (open
and closed) string partition functions are identical to those in ~\cite{FOE}. 
Like in~\cite{FOE}, we label the Ishibashi states of~\R{torus} by
pairs $(m,J)$ that obey
\begin{eqnarray}
m & = & Jm \;\;\;,\\
Q_K(m) + X(K,J) &= & 0 \mod 1 
\end{eqnarray}
for all $K \in \cH$. 
The boundary labels $[a,\psi_a]$ are $\cH$-orbits $[a]$ of a chiral sector $a$. We also need a
 boundary degeneracy label
 $\psi_a$. It is a discrete group character of the central stabilizer $\cC_a$ (see below).
The boundary states are determined by boundary coefficients. In the simple current case, these are 
\beq \label{eq:R}
R_{[a,\psi_a](m,J)} = \sqrt{\frac{|\cH|}{|\cC_a||\cS_a|}}
 \psi_a^*(J) S^J_{am}
\eeq
The {\em fixed point resolution matrix} $S^J$, whose rows and columns are labelled by fixed points $a,m$ of $J$,
implements a modular $S$-transformation
 on the torus with $J$ inserted.
 It is unitary and obeys \cite{FSS}
\beq
S^J_{Ki,j} = F_i(K,J) e^{2\pi\ii Q_K(j)} S^J_{ij} \;\;\; .
\eeq
The phase $F$ is called the simple current twist. We can now define the central stabilizer as
\beq
\cC_a = \{J \in \cS_a | F_a(K,J)e^{2\pi\ii X(K,J)} = 1 \;\; {\rm for} \;\; {\rm all} \;\; K \in \cS_a\} .
\eeq
For more details we refer to~\cite{thesis,FOE}.
In contrast to \cite{FOE} these boundary coefficients are the same for all orientifold choices, and
are in fact also valid for oriented strings and non-symmetric modular invariants. They can be
used to compute oriented annulus coefficients, defined
as
\beq  \label{eq:NIMREP}
A^{i~~[b,\psi_b]}_{~[a,\psi_a]} = \sum_{m,J} \frac{ S^i_{~m}R_{[a,\psi_a](m,J)}( R_{[b,\psi_b](m,J)})^*}{S_{0m}}
\eeq
Now we have to introduce orientifold choices, and to do so we restrict ourselves to symmetric modular invariants.
This implies that we only consider symmetric matrices $X$. The orientifold choice enters into the
formalism in two ways, namely through the crosscap coefficients and the definition of the unoriented Annulus. 
The allowed choices are as follows.  One must select
\begin{enumerate}
\item  A {\em Klein bottle current} $K$~\cite{klein}. This can be any simple current
 of $\cA$ that is local with all order two currents in $\cH$. Only odd currents outside $\cH$ can give
 spectra that are inequivalent to those with $K=0$. See~\cite{thesis} for details.
\item A set of phases $\b(J)$ for all $J\in \cH$ that satisfy
\beq 
\b(J) \b(J') = \b(JJ') e^{2\pi\ii X(J,J')}  \;\;\; , J,J' \in \cH
\eeq
with $\b(0) = e^{\pi\ii h_K}$.
\end{enumerate} 
In the latter case the freedom is due to that fact that for every even cyclic factor in $\cH$ a sign remains
undetermined by this condition. The crosscap coefficient of the orientifold $(\cA_{st},\cH,X,K,\b)$ is 
\beq \label{eq:U}
U^{\Omega}_{(m,J)} = \frac{1}{\sqrt{|\cH |}} \sum_{L\in \cH}
\s(L) P_{LK,m}\delta_{J,0}  \;\;\; , \s(L) := \b(L)e^{\pi{\rm i} [h_{LK}- h_K]} \;\;\; .
\eeq 
Here $\Omega$ is a generic notation for the possible orientifold choices.
One can show that the $\s(L)$ are signs. The matrix $P=\sqrt{T}ST^2S\sqrt{T}$~\cite{P}. 
The unoriented annulus is given by
\beq
A^{\Omega,i}_{[a,\psi_a][b,\psi_b]} = \sum_{m,J,J'} \frac{ S^i_{~m}R_{[a,\psi_a](m,J)} g^{\Omega,m}_{J,J'} R_{[b,\psi_b](m,J')} }{S_{0m}}
\eeq
The {\em Ishibashi metric} $g^{\Omega,m}$ is defined as
\beq
g^{\Omega,m}_{J,J'} =  \frac{S_{m0}}{S_{mK}} \beta(J) \delta_{J',J^c}
\eeq  
The Moebius and Klein bottle amplitude follow from
\beq
M^i_{[a,\psi_a]} = \sum_{m,J,J'} \frac{ P^i_{~m}R_{[a,\psi_a](m,J)} g^{\Omega,m}_{J,J'} U_{(m,J')} }{S_{0m}}
\eeq
\beq
K^i = \sum_{m,J,J'} \frac{ S^i_{~m}U_{(m,J)} g^{\Omega,m}_{J,J'} U_{(m,J')} }{S_{0m}}
\eeq  
The unoriented annuli for the various choices of $\Omega$ can all be derived from the unique oriented annulus 
\R{NIMREP}
by matrix multiplication with the boundary conjugation matrix $A^{\Omega,0}$ that maps a brane $[a,\psi_a]$ to
its orientifold image $[a,\psi_a]^c$.

In~\cite{thesis} it is shown that the spectrum is positive and integral. 
Note that this only determines
the boundary and crosscap coefficients up to a common $(m,J)$ dependent sign.
Presumably these signs can be determined by solving the sewing constraints, but fortunately
they are not relevant for our purposes. In addition integrality is unaffected by the overall
sign of the crosscap coefficients. 
Recently in \cite{Fuchs:2004dz} it was demonstrated that the resulting boundary CFTs are
consistent on all orientable surfaces (work on non-orientable surfaces is in progress).

\subsection{Tadpole Cancellation}  

It is our goal to construct stable, finite, supersymmetric four-dimensional string theories. In other
words, we will insist on the cancellation of all tadpoles due to both NS-NS and R-R massless scalars.
In chiral models, this implies in particular the cancellation of the cubic part of
the gauge anomalies~\cite{BiM}.
The tadpole cancellation conditions are equations for the
Chan-Paton multiplicities $N_{[a,\psi_a]}$, and take the form
\beq
\sum_{[a,\psi_a]} N_{[a,\psi_a]} R_{[a,\psi_a](m,J)} = 4 \epsilon \eta_m U_{(m,J)}
\eeq
for all Ishibashi labels $(m,J)$ that correspond to massless closed strings in~\R{torus}.
Here $\eta_m=1$ for $m=0$, the vacuum, and $\eta_m=-1$ otherwise, 
and $\epsilon$ is the overall crosscap sign. It is fixed by
the dilaton tadpole condition.
Note that the aforementioned $(m,J)$ dependent signs cancel in the tadpole
equations as well.

In principle one could proceed by solving these equations by computer. This is indeed possible
in the six-dimensional case, where we have obtained the complete solution for all orientifolds of
all simple current invariants of all $c=6$ tensor products of $N=2$ minimal models (see
\cite{comments} for some special cases.) This is a very useful test of the entire formalism, since
anomaly cancellation in six dimensions is a far more powerful constraint then it is in four dimensions.

This method is not feasible in four dimensions, because the number of variables vastly outnumbers
the number of conditions. In one of the 168 cases we have been able to do this (for all simple current
invariants and orientifolds), namely $(1,3,3,4,8)$ (which has only 260 primaries, and a chiral spectrum in the Cardy case).

In all other cases we proceed as follows. First we determine a subset of boundaries that produces
a desired spectrum, for example the standard model or some of its extensions. Unless one is extremely
lucky this set of boundaries and CP multiplicies will not satisfy the tadpole conditions by itself. 
Therefore we allow additional ``hidden" branes. Of course there might be open strings stretching
between the  standard model and the hidden branes. We will allow such states provided that they
are not chiral. The details will be discussed now.

\subsection{Chiral Spectrum} \label{sec-spectra}

There are many conceivable intersecting brane realizations of the Standard Model, but
we will aim here for the simplest kind, and in particular
the smallest number of branes. We will require that all standard model particles
come from strings between different branes (bi-fundamentals), and that baryon and lepton number are conserved
perturbatively. This leads almost inevitably to models with four stacks of branes. 
Following \cite{IbMR0105} we will label them $a,b,c$ and $d$. 
The color gauge group $SU(3)$ is associated
with brane $a$ and its orientifold image $a^c$, which must produce a Chan-Paton group $U(3)_a$.
The weak gauge group $SU(2)$ is associated with brane $b$; this
group can either be $U(2)$ or $Sp(2)$; in the latter case $b=b^c$. Branes $c$ and $c^c$ have
a $U(1)$ CP-group, as do $d$ and $d^c$. Baryon number is related to $U(1)_a$,
and lepton number to $U(1)_d$. 
The standard model $Y$-charge is given by $\frac16 Q_a - \frac12 Q_c - \frac12 Q_d$.  

In table I we summarize all the massless particles with standard model gauge representations 
that can in principle occur with this brane
configuration. We treat
all particles as left-handed.
The brane representations are denoted as $V$ for vectors, ``Adj" for adjoint,
``A" for antisymmetric tensor and ``S" for symmetric tensor, and a $*$ denotes complex conjugation.
The sections of the table denote respectively standard model particles, Higgses,
exotics which respect standard model charge quantization, exotics that do not, and hidden matter. In the last column we indicate the $SU(3)\times SU(2)\times U(1)$ quantum numbers of the
particles. For representations that do not occur in the standard model we specify,
as a subscript, three times baryon number, and lepton number.

To every row $i=1,...,29$ we associate two non-negative integer multiplicities, $M_i$ and $\bM_i$, where
the former is the multiplicity of the particle as shown, and the latter is that of
its complex conjugate (of the full representation, including the hidden sector).
 Of course $M_{16}=\bM_{16},M_{19}=\bM_{19}$ and $M_{22}=\bM_{22}$.
There is a redundancy in the table if the weak group is $Sp(2)$, and hence in that case
we can set $M_2=\bM_2=M_6=\bM_6=M_{16}=0$ without loss of generality. In the Higgs
sector, standard model anomaly cancellation requires that there be an equal number
of representations $H_1=(1,2,-\frac12)$ and $H_2=(1,2,\frac12)$.  Hence $M_9+\bM_{10}=M_{10}+\bM_9\equiv M_H$.
In the case of a weak group $Sp(2)$ the parameters $\bM_9$ and $\bM_{10}$ are redundant, and hence in
that case $M_9=M_{10}=M_H$. 

Our requirements on the spectrum are as follows. Define $\Delta_i=M_i-\bM_i$. Then we require
that $\Delta_1+\Delta_2=\Delta_3=\Delta_4=\Delta_5+\Delta_6=\Delta_7=\Delta_8=3$. We impose
no restriction on $\Delta_9$ and $\Delta_{10}$, except that they should be equal as explained above.
Note that for $Sp(2)$ $\Delta_9=\Delta_{10}=0$.
All $\Delta_i$ of the exotic representations, $i=11,\ldots,28$ are required to vanish. 
This implies that mass
terms are allowed for all exotic representations without breaking the original Chan-Paton group. We
are implicitly assuming that such mass terms are indeed generated, so that only the chiral part of the
spectrum is accessible with current experiments. Such mass terms may indeed be generated if one
moves away from the rational point in the moduli space. We do allow $\Delta_{29} \not=0$, \ie\
the hidden sector may be chiral.

Exotics
not respecting charge quantization (nrs. 25 to 28)
may occur due to strings stretching between the standard model and the hidden branes.
Indeed, the corresponding
color singlets always have half-integer electric charge, and hence at least one of them would be
stable. There is a variety of ways around this. First of all, these particles may be sufficiently
massive and rare to have escaped attention so far; secondly, they may be confined to integer charge
hadron like particles by a gauge group from the hidden sector; thirdly, they may simply be absent from the spectrum, and
finally the entire hidden gauge sector may be absent, so that they cannot occur at all. 
Indeed, we found examples of the latter two possibilities, as well as cases where all these strings end
on a $U(2)$ or $U(4)$ hidden brane, which are plausible candidates for the second option. On the other hand,
one could regard half-integer charge particles as a fairly generic (though not fully general) prediction
of this class of models.

In addition to the particles listed in the table there may exist particles that belong
entirely to the hidden sector. We impose no further conditions on this matter. It may in fact
even by chiral, although in most cases we have found it is not.

\vskip .3truecm
\renewcommand{\arraystretch}{1.2}
\begin{table}[h!]
\footnotesize
\begin{center}
~~~~~~~~~~~~~~~~\begin{tabular}{|l|c|c|c|c|c|} \hline
nr. &  $U(3)_a$ & Weak & $U(1)_c$ & $U(1)_d$  & massless particle  \\ \hline \hline
1 & $V$ & $V$ &  0 & 0  & $(u,d)$  \\
2 & $V$ & $V^*$ &  0 & 0  & $(u,d)$  \\
3 & $V^*$ & 0 &  $V$ & 0  & $u^c$  \\
4 & $V^*$ & 0 &  $V^*$ & 0  & $d^c$  \\
5 &  0 & $V$ &  0 & $V$  & $(\nu,e^{-})$  \\
6 & 0  &$V^*$ &  0 & $V$  & $(\nu,e^{-})$  \\
7 & 0 & 0 &  $V$ & $V^*$  & $\nu^c$  \\
8 & 0 & 0 &  $V^*$ & $V^*$  & $e^{+}$  \\  \hline
9 & 0 & $V$ &  $V$ & 0  & $H_1$  \\
10 & 0 & $V$ &  $V^*$ & 0  & $H_2$  \\ \hline
11 & $V$ & 0 &  0 & $V$  & $(3,1,-\frac13)_{1,1}$  \\
12 & $V$ & 0 &  0 & $V^*$ &  $(3,1,\frac23)_{1,-1}$  \\
13 & Adj & 0 &  0 & 0 &  $(8,1,0)_{0,0}+(1,1,0)_{0,0}$  \\
14 & A & 0 &  0 & 0 &  $(3^*,1,\frac13)_{2,0}$  \\
15 & S & 0 &  0 & 0 &  $(6,1,\frac13)_{2,0}$  \\
16 & 0 & Adj &  0 & 0 &  $(1,3,0)_{0,0}+(1,1,0)_{0,0}$  \\
17 & 0 & A &  0 & 0 &  $(1,1,0)_{0,0}$  \\
18 & 0 & S &  0 & 0  &  $(1,3,0)_{0,0}$  \\
19 & 0 & 0 &  Adj & 0 &  $(1,1,0)_{0,0}$  \\
20 & 0 & 0 &  A & 0 &  ---  \\
21 & 0 & 0 &  S & 0 &  $(1,1,-1)_{0,0}$  \\
22 & 0 & 0 &  0 & Adj &  $(1,1,0)_{0,0}$  \\
23 & 0 & 0 &  0 & A &  ---  \\
24 & 0 & 0 &  0 & S &  $(1,1,-1)_{0,2}$  \\ \hline
25 & $V$ & 0 &  0 & 0 &  $(3,1,\frac16)_{1,0}$  \\
26 & 0 & $V$ &  0 & 0 &  $(1,2,0)_{0,0}$  \\
27 & 0 & 0 &  $V$ & 0 &  $(1,1,-\frac12)_{0,0}$  \\
28 & 0 & 0 &  0 & $V$ &  $(1,1,-\frac12)_{0,1}$  \\ \hline
29 & 0 & 0 &  0 & $0$ &  $(1,1,0)_{0,0}$ \\ \hline
\end{tabular}
\caption{List of standard model representations that can appear, and their labelling.}
\end{center}
\end{table}
\renewcommand{\arraystretch}{1.0}

\section{Results} \label{sec-models}

In practice, most of the 168 $c=9$ tensor combinations are
accessible by our methods. Indeed, we have been able to analyse 160 combinations under more
restrictive conditions than those formulated above.
The type of spectra described in the previous section have so far been searched for in
66 tensor products. In 48 of them the required standard model brane
configuration did not occur for any modular invariant and orientifold choice.
In 13 other cases they did occur, but it was possible to show that the tadpole equations have no solution.

In five cases we did find spectra that satisfy our criteria. At this moment,
we have found such examples for the models listed in table 2. The first column
gives the levels of the minimal models. In order to identify the MIPF, in the
second column we list the hodge numbers and the number of gauge singlets in the
corresponding  heterotic string spectrum, for comparison with the tables of \cite{Schellekens:1989wx,Fuchs:1990yv}.
In the third column we give the number of Ishibashi labels, or equivalently, the number
of boundaries. In column four we specify for how many orientifold choices we have found
solutions to all the tadpole equations satisfying our criteria. For the benefit of the reader
we give also the result for the Cardy case, although no solutions were found. 
It is amusing to note that in all cases with
solutions $h_{21} < h_{11}$.

We emphasize
that this table is by no means complete. In fact, we expect there to exist a large, possibly
astronomical number of additional solutions. We have only partly explored the remaining 102 tensor products.
In some of them the tadpole
conditions are unsolvable because the
number of candidate hidden branes is simply too large.

For all of the cases listed in the table we find many standard model brane configurations,
and for many of them a large number of ways of adding hidden sector branes that saturate
the tadpole conditions. For example, for tensor $(6,6,6,6)$, invariant $(3,59,223)$ we have
so far identified more than 6000 distinct solutions, without even carefully distinguishing 
all features of the hidden sector, and by only considering the minimal number of
hidden branes. These spectra differ in at least one of the integers $M_i$ defined above, and/or
in the hidden sector gauge group.

We are still analyzing this enormous set of solutions, and just give here some fairly
randomly chosen examples. We only present the open sector;
in addition there are of course massless particles from the closed sector. All examples we
discuss below have a non-chiral hidden sector.

To specify a model without extra branes it is sufficient to give the integers $M_i$. A typical
example with an Sp(2) weak gauge group has the following spectrum:
quarks and leptons: $(M_1,M_3,M_4,M_5,M_7,M_8)=(3,3,3,4,6,6)$, Adj: $(2,0,5,18)$, A: $(1,4,2,9)$, S:$(1,1,4,9)$
(these are the number of these representations in each of the four factors)
lepto-quarks: $(M_{11}, M_{12})=(6,0)$, Higgs: $M_9=M_{10}=3$. Note that, for example, $M_5=4$ means that
there are four left-handed lepton doublets, and one right-handed one, since the chiral part of  the spectrum
is always fixed in the way discussed before. For the same reason the (anti)-symmetric tensors and lepto-quarks
are always non-chiral, \ie\ $\bar M_i=\bM_i$, for $i=11 \ldots 24$.
This model has three copies of Higgs bosons in the
standard MSSM representation $(1,2,\frac12)+(1,2,-\frac12)$.
If there are extra hidden branes there is usually a large number of options.
A simple example with just a single
extra $U(1)$-brane and no SM-hidden bifundamentals of types $25 \ldots 28$
has the following characteristics:  $(M_1,M_3,M_4,M_5,M_7,M_8)=(5,5,5,7,5,5)$,
two times $H_1+H_2$, two lepto-quarks of each charge $(M_{11}, M_{12})=(2,2)$, and the following
multiplicities for the rank-2 fields in the five groups: Adj: (2,0,2,8,1); S: (1,1,1,5,2); A: (2,1,2,4,0). 

Another example has no mirror quarks and leptons 
at all, but two extra CP groups $U(9)\times U(1)$ coupling to non-chiral matter of types $25\ldots 29$.
We will not present here
the rather large numbers of non-chiral matter in (anti)-symmetric tensors and adjoints,
which appears to be a generic feature of these spectra.

Among the models with a weak symmetry group $U(2)$ there are cases 
with 
$(M_1,\ldots,M_8)$ \newline $ = (1,2,3,3,3,3,3,3)$, 
but unfortunately
no Higgs at all. In this case there are no mirror quarks and leptons, so that the quark/lepton spectrum
is exactly that described in \cite{IbMR0105}. There are additional gauge groups and 
non-chiral rank-2 fields.
There are other $U(2)$ cases with a few mirror quarks or leptons and some Higgs bosons with vanishing brane chirality
($\Delta_9=\Delta_{10}=0$). We will not give more examples here; at the moment an important challenge is how
to select the most attractive ones from the huge list.

\renewcommand{\arraystretch}{1.2}
\begin{table}[h!]
\footnotesize
\begin{center}
~~~~~~~~~~~~~~~~\begin{tabular}{|l|c|c|c|c|c|} \hline
tensor &  $(h_{21},h_{11},S) $ & Boundaries & Orientifolds  \\ \hline \hline
(6,6,6,6) &(149,1,503) & 9632 &  ---  \\
          &(5,69,267) & 400 & 2 \\
          &(9,41,211) & 800 & 2 \\
          &(3,59,223) & 368 & 4 \\
          &(5,37,203) & 368 & 4 \\
          & (3,43,207) & 400 & 1 \\
          & (17,25,203) & 1136 & 1 \\ \hline
(3,8,8,8) & (145,1,495) & 9200 & --- \\
          & (11,47,283) & 880 & 1 \\ \hline
(4,6,6,10) & (66,6,281) & 1540 & --- \\
          & (14,38,229) & 416 & 1 \\ \hline
(4,4,10,10) & (128,2,443) & 7200 & --- \\
          & (10,64,229) & 406 & 1 \\ \hline
(2,5,12,26) & (116,8,453) & 6006 & --- \\
          & (23,59,327) & 780 & 1 \\
          & (23,59,327) & 858 & 1 \\ \hline
\end{tabular}
\caption{Modular invariants for which chiral SSM were found so far. 
The last column gives the number of distinct orientifold choices which have solutions.}
\end{center}
\end{table}
\renewcommand{\arraystretch}{1.0}

\section{Conclusion}

The main goal of this paper is to point out that large numbers of vacua with just 
the chiral standard model spectra can be obtained from orientifolds of
non-toric Calabi-Yau compactifications. Since these CFTs correspond to special points in
a multi-dimensional moduli space, we focused here on features
that are most robust under changes of the moduli: the chiral spectrum.
Even with the very limited search we have done
so far the number of solutions is enormous, and
more detailed phenomenological input
would be needed to reduce this to
a more managable set. 

Most examples found so far have an $Sp(2)$ weak gauge group and a large quantity of
non-chiral additional matter. We did find examples with a $U(2)$ weak gauge group, but so far
they all had Higgses with zero brane chirality. There are examples without hidden branes, without
any mirror quarks and leptons, without any half-integer charge exotic matter (even though there is a
hidden sector) and examples with hidden sectors that are capable of confining the half-integer charges
to integer charges.

We have limited ourselves here to two simple and
attractive brane realizations of the standard model. Still more
solutions would undoubtedly be found if we allow realizations of the
standard model gauge group in larger Chan-Paton groups. 
On the other hand,
with better {\it a priori} constraints
one could do a dedicated search for models with such desirable features. 

The intrinsic limitation of RCFT methods is that one is working on a given
point in moduli space.
There are many phenomenological issues that could be discussed, but for many
of them this restriction is important. For example, three point couplings 
(in particular fermion-Higgs couplings) are computable in principle in RCFT,
but are also modulus dependent. Therefore we see these results primarily as 
a guide to interesting regions in CY-moduli space.

One issue that can be discussed in RCFT and may remain valid beyond it is the
mass of $U(1)$ gauge bosons. In general, $U(1)$ mixed anomalies are of the form
$\Tr F_a \Tr (F_b)^2 $. They are cancelled by a Green-Schwarz mechanism
involving couplings of RR two-form fields to
$\Tr (F_b)^2$  and  $\Tr F_a $. The latter kind of couplings give masses to 
$U(1)$ gauge bosons.  They must be present for anomalous $U(1)$'s, but may
also be present for anomaly-free ones~\cite{IbMR0105}.
In our case Baryon and Lepton number are anomalous, but
$B-L$ and of course $Y$ are not. 
We have worked out the coupling of these
gauge bosons to the two-form fields and found that in a surprisingly
large number of cases all such
couplings vanish, for both $B-L$ and $Y$. This includes spectra without hidden branes.
This implies that both gauge bosons
have zero mass, and that a mass for the $B-L$ gauge boson has to be generated
by some other mechanism in order for these spectra to be acceptable.
There are also cases where both the $B-L$ and the $Y$ gauge boson have non-zero mass,
or or only one of the two.
In particular we have examples with vanishing $Y$-mass, and non-vanishing
$B-L$ gauge boson mass. Some of these examples have no mirror quarks and leptons,
but they do have additional gauge groups and non-chiral exotic matter.

We have not yet done a complete analysis of all abelian gauge boson masses in all models
we have found so far, and in addition we expect a large number of additional cases to appear
when we explore the remaining tensor products. The results of a more complete
survey will be presented in a forthcoming publication.

\bigskip

{\bf Acknowledgements:}

L.R.H. is grateful to I. Demasure, B. Janssen, J. Gheerardyn,
T. van Proeyen, W. Troost, A. Uranga and especially T. Ott for discussions.
A.N.S. wishes to thank B. Gato-Rivera for discussions and IMAFF-CSIC,
Madrid, where part of this work was done, for hospitality and computing
support. The
work of A.N.S. has been performed as part of the program
FP 52 of the Foundation for Fundamental Research of Matter (FOM), and
the work of T.P.T.D. and A.N.S. has been performed as part of the
program FP 57 of FOM.
The work A.N.S. has been partially
supported by funding of the Spanish ``Ministerio de
Ciencia y Tecnolog\'\i a", Project BFM2002-03610. We thank I. Runkel for comments 
on an earlier version of this letter.

\bibliography{REFS}

\providecommand{\href}[2]{#2}\begingroup\raggedright\begin{thebibliography}{10}

\bibitem{BlGKL0007}
R.~Blumenhagen, L.~Goerlich, B.~Kors  and D.~Lust, \emph{Noncommutative
  compactifications of type I strings on tori with magnetic background flux},
  JHEP {\bf 10} (2000) 006,
\href{http://www.arXiv.org/abs/hep-th/0007024}{{\tt hep-th/0007024}}

\bibitem{AlFIRU0011}
G.~Aldazabal, S.~Franco, L.~E. Ibanez, R.~Rabadan  and A.~M. Uranga, \emph{D =
  4 chiral string compactifications from intersecting branes}, J. Math. Phys.
  {\bf 42} (2001) 3103--3126,
\href{http://www.arXiv.org/abs/hep-th/0011073}{{\tt hep-th/0011073}}

\bibitem{IbMR0105}
L.~E. Ibanez, F.~Marchesano  and R.~Rabadan, \emph{Getting just the standard
  model at intersecting branes}, JHEP {\bf 11} (2001) 002,
\href{http://www.arXiv.org/abs/hep-th/0105155}{{\tt hep-th/0105155}}

\bibitem{BlKLO0107}
R.~Blumenhagen, B.~Kors, D.~Lust  and T.~Ott, \emph{The standard model from
  stable intersecting brane world orbifolds}, Nucl. Phys. {\bf B616} (2001)
  3--33,
\href{http://www.arXiv.org/abs/hep-th/0107138}{{\tt hep-th/0107138}}

\bibitem{CvSU0107}
M.~Cvetic, G.~Shiu  and A.~M. Uranga, \emph{Chiral four-dimensional N = 1
  supersymmetric type IIA orientifolds from intersecting D6-branes}, Nucl.
  Phys. {\bf B615} (2001) 3--32,
\href{http://www.arXiv.org/abs/hep-th/0107166}{{\tt hep-th/0107166}}

\bibitem{BaKL0108}
D.~Bailin, G.~V. Kraniotis  and A.~Love, \emph{Standard-like models from
  intersecting D4-branes}, Phys. Lett. {\bf B530} (2002) 202--209,
\href{http://www.arXiv.org/abs/hep-th/0108131}{{\tt hep-th/0108131}}

\bibitem{BlBKL0206}
R.~Blumenhagen, V.~Braun, B.~Kors  and D.~Lust, \emph{Orientifolds of K3 and
  Calabi-Yau manifolds with intersecting D-branes}, JHEP {\bf 07} (2002) 026,
\href{http://www.arXiv.org/abs/hep-th/0206038}{{\tt hep-th/0206038}}

\bibitem{Ko0207}
C.~Kokorelis, \emph{Exact standard model structures from intersecting D5-
  branes}, Nucl. Phys. {\bf B677} (2004) 115--163,
\href{http://www.arXiv.org/abs/hep-th/0207234}{{\tt hep-th/0207234}}

\bibitem{BlGO0211}
R.~Blumenhagen, L.~Gorlich  and T.~Ott, \emph{Supersymmetric intersecting
  branes on the type IIA T**6/Z(4) orientifold}, JHEP {\bf 01} (2003) 021,
\href{http://www.arXiv.org/abs/hep-th/0211059}{{\tt hep-th/0211059}}

\bibitem{CrIM0302}
D.~Cremades, L.~E. Ibanez  and F.~Marchesano, \emph{Yukawa couplings in
  intersecting D-brane models}, JHEP {\bf 07} (2003) 038,
\href{http://www.arXiv.org/abs/hep-th/0302105}{{\tt hep-th/0302105}}

\bibitem{H0303}
G.~Honecker, \emph{Chiral supersymmetric models on an orientifold of Z(4) x
  Z(2) with intersecting D6-branes}, Nucl. Phys. {\bf B666} (2003) 175--196,
\href{http://www.arXiv.org/abs/hep-th/0303015}{{\tt hep-th/0303015}}

\bibitem{CaU0303}
J.~F.~G. Cascales and A.~M. Uranga, \emph{Chiral 4d N = 1 string vacua with
  D-branes and NSNS and RR fluxes}, JHEP {\bf 05} (2003) 011,
\href{http://www.arXiv.org/abs/hep-th/0303024}{{\tt hep-th/0303024}}

\bibitem{LaP0305}
M.~Larosa and G.~Pradisi, \emph{Magnetized four-dimensional Z(2) x Z(2)
  orientifolds}, Nucl. Phys. {\bf B667} (2003) 261--309,
\href{http://www.arXiv.org/abs/hep-th/0305224}{{\tt hep-th/0305224}}

\bibitem{CvLL0403}
M.~Cvetic, T.~Li  and T.~Liu, \emph{Supersymmetric Pati-Salam models from
  intersecting D6- branes: A road to the standard model},
\href{http://www.arXiv.org/abs/hep-th/0403061}{{\tt hep-th/0403061}}

\bibitem{Ur0301}
A.~M. Uranga, \emph{Chiral four-dimensional string compactifications with
  intersecting D-branes}, Class. Quant. Grav. {\bf 20} (2003) S373--S394,
\href{http://www.arXiv.org/abs/hep-th/0301032}{{\tt hep-th/0301032}}

\bibitem{Ot0309}
T.~Ott, \emph{Aspects of stability and phenomenology in type IIA orientifolds
  with intersecting D6-branes}, Fortsch. Phys. {\bf 52} (2004) 28--137,
\href{http://www.arXiv.org/abs/hep-th/0309107}{{\tt hep-th/0309107}}

\bibitem{K0310}
E.~Kiritsis, \emph{D-branes in standard model building, gravity and cosmology},
  Fortsch. Phys. {\bf 52} (2004) 200--263,
\href{http://www.arXiv.org/abs/hep-th/0310001}{{\tt hep-th/0310001}}

\bibitem{Lu0401}
D.~Lust, \emph{Intersecting brane worlds: A path to the standard model?},
\href{http://www.arXiv.org/abs/hep-th/0401156}{{\tt hep-th/0401156}}

\bibitem{CvPS0212}
M.~Cvetic, I.~Papadimitriou  and G.~Shiu, \emph{Supersymmetric three family
  SU(5) grand unified models from type IIA orientifolds with intersecting
  D6-branes}, Nucl. Phys. {\bf B659} (2003) 193--223,
\href{http://www.arXiv.org/abs/hep-th/0212177}{{\tt hep-th/0212177}}

\bibitem{CvP0303}
M.~Cvetic and I.~Papadimitriou, \emph{More supersymmetric standard-like models
  from intersecting D6-branes on type IIA orientifolds}, Phys. Rev. {\bf D67}
  (2003) 126006,
\href{http://www.arXiv.org/abs/hep-th/0303197}{{\tt hep-th/0303197}}

\bibitem{BiSa}
M.~Bianchi and A.~Sagnotti, \emph{On the systematics of open string theories},
  Phys. Lett. {\bf B247} (1990)
517--524

\bibitem{GiP}
E.~G. Gimon and J.~Polchinski, \emph{Consistency Conditions for Orientifolds
  and D-Manifolds}, Phys. Rev. {\bf D54} (1996) 1667--1676,
\href{http://www.arXiv.org/abs/hep-th/9601038}{{\tt hep-th/9601038}}

\bibitem{Cardy}
J.~L. Cardy, \emph{Boundary conditions, fusion rules and the verlinde formula},
  Nucl. Phys. {\bf B324} (1989)
581

\bibitem{Sa}
A.~Sagnotti, \emph{Open strings and their symmetry groups}, in Non-Perturbative
  Quantum Field Theory, eds. G. Mack et al. (1987)
\href{http://www.arXiv.org/abs/hep-th/0208020}{{\tt hep-th/0208020}}

\bibitem{BlW9806}
R.~Blumenhagen and A.~Wisskirchen, \emph{Spectra of 4D, N = 1 type I string
  vacua on non-toroidal CY threefolds}, Phys. Lett. {\bf B438} (1998) 52--60,
\href{http://www.arXiv.org/abs/hep-th/9806131}{{\tt hep-th/9806131}}

\bibitem{GoM0306}
S.~Govindarajan and J.~Majumder, \emph{Crosscaps in Gepner models and type IIA
  orientifolds}, JHEP {\bf 02} (2004) 026,
\href{http://www.arXiv.org/abs/hep-th/0306257}{{\tt hep-th/0306257}}

\bibitem{AlALN0307}
G.~Aldazabal, E.~C. Andres, M.~Leston  and C.~Nunez, \emph{Type IIB
  orientifolds on Gepner points}, JHEP {\bf 09} (2003) 067,
\href{http://www.arXiv.org/abs/hep-th/0307183}{{\tt hep-th/0307183}}

\bibitem{Bl0310}
R.~Blumenhagen, \emph{Supersymmetric orientifolds of Gepner models}, JHEP {\bf
  11} (2003) 055,
\href{http://www.arXiv.org/abs/hep-th/0310244}{{\tt hep-th/0310244}}

\bibitem{BrHHW0401}
I.~Brunner, K.~Hori, K.~Hosomichi  and J.~Walcher, \emph{Orientifolds of Gepner
  models},
\href{http://www.arXiv.org/abs/hep-th/0401137}{{\tt hep-th/0401137}}

\bibitem{BlW0401}
R.~Blumenhagen and T.~Weigand, \emph{Chiral supersymmetric Gepner model
  orientifolds},
\href{http://www.arXiv.org/abs/hep-th/0401148}{{\tt hep-th/0401148}}

\bibitem{FOE}
J.~Fuchs, L.~R. Huiszoon, A.~N. Schellekens, C.~Schweigert  and J.~Walcher,
  \emph{Boundaries, crosscaps and simple currents}, Phys. Lett. {\bf B495}
  (2000) 427--434,
\href{http://www.arXiv.org/abs/hep-th/0007174}{{\tt hep-th/0007174}}

\bibitem{Po}
J.~Polchinski, \emph{Dirichlet-Branes and Ramond-Ramond Charges}, Phys. Rev.
  Lett. {\bf 75} (1995) 4724--4727,
\href{http://www.arXiv.org/abs/hep-th/9510017}{{\tt hep-th/9510017}}

\bibitem{SaS}
A.~Sagnotti and Y.~S. Stanev, \emph{Open descendants in conformal field
  theory}, Fortsch. Phys. {\bf 44} (1996) 585--596,
\href{http://www.arXiv.org/abs/hep-th/9605042}{{\tt hep-th/9605042}}

\bibitem{P}
G.~Pradisi, A.~Sagnotti  and Y.~S. Stanev, \emph{Planar duality in SU(2) WZW
  models}, Phys. Lett. {\bf B354} (1995) 279--286,
\href{http://www.arXiv.org/abs/hep-th/9503207}{{\tt hep-th/9503207}}

\bibitem{SBB1}
J.~Fuchs and C.~Schweigert, \emph{Symmetry breaking boundaries. I: General
  theory}, Nucl. Phys. {\bf B558} (1999) 419--483,
\href{http://www.arXiv.org/abs/hep-th/9902132}{{\tt hep-th/9902132}}

\bibitem{SBB2}
J.~Fuchs and C.~Schweigert, \emph{Symmetry breaking boundaries. II: More
  structures, examples}, Nucl. Phys. {\bf B568} (2000) 543--593,
\href{http://www.arXiv.org/abs/hep-th/9908025}{{\tt hep-th/9908025}}

\bibitem{klein}
L.~R. Huiszoon, A.~N. Schellekens  and N.~Sousa, \emph{Klein bottles and simple
  currents}, Phys. Lett. {\bf B470} (1999) 95--102,
\href{http://www.arXiv.org/abs/hep-th/9909114}{{\tt hep-th/9909114}}

\bibitem{GaS92}
B.~Gato-Rivera and A.~N. Schellekens, \emph{Complete classification of simple
  current modular invariants for (Z(p))**k}, Commun. Math. Phys. {\bf 145}
  (1992)
85--122

\bibitem{KrS}
M.~Kreuzer and A.~N. Schellekens, \emph{Simple currents versus orbifolds with
  discrete torsion: A Complete classification}, Nucl. Phys. {\bf B411} (1994)
  97--121,
\href{http://www.arXiv.org/abs/hep-th/9306145}{{\tt hep-th/9306145}}

\bibitem{Gepner:1988qi}
D.~Gepner, \emph{Space-Time Supersymmetry In Compactified String Theory And
  Superconformal Models}, Nucl. Phys. {\bf B296} (1988)
757

\bibitem{LeSW}
W.~Lerche, A.~N. Schellekens  and N.~P. Warner, \emph{Lattices and strings},
  Phys. Rept. {\bf 177} (1989)
1

\bibitem{FSW}
J.~Fuchs, C.~Schweigert  and J.~Walcher, \emph{Projections in string theory and
  boundary states for Gepner models}, Nucl. Phys. {\bf B588} (2000) 110--148,
\href{http://www.arXiv.org/abs/hep-th/0003298}{{\tt hep-th/0003298}}

\bibitem{HS}
L.~R. Huiszoon and K.~Schalm, \emph{BPS orientifold planes from crosscap states
  in Calabi-Yau compactifications}, JHEP {\bf 11} (2003) 019,
\href{http://www.arXiv.org/abs/hep-th/0306091}{{\tt hep-th/0306091}}

\bibitem{Lutken:1988zj}
C.~A. Lutken and G.~G. Ross, \emph{Symmetries And Couplings In Heterotic
  Superconformal Field Theories}, Phys. Lett. {\bf B214} (1988)
357

\bibitem{Lynker:fs}
M.~Lynker and R.~Schimmrigk, \emph{On The Spectrum Of (2,2) Compactification Of
  The Heterotic String On Conformal Field Theories}, Phys. Lett. {\bf B215}
  (1988)
681

\bibitem{Schellekens:1989wx}
A.~N. Schellekens and S.~Yankielowicz, \emph{New Modular Invariants For N=2
  Tensor Products And Four-Dimensional Strings}, Nucl. Phys. {\bf B330} (1990)
103

\bibitem{Fuchs:1989pt}
J.~Fuchs, A.~Klemm, C.~Scheich  and M.~G. Schmidt, \emph{Gepner Models With
  Arbitrary Affine Invariants And The Associated Calabi-Yau Spaces}, Phys.
  Lett. {\bf B232} (1989)
317

\bibitem{Fuchs:1990yv}
J.~Fuchs, A.~Klemm, C.~Scheich  and M.~G. Schmidt, \emph{Spectra And Symmetries
  Of Gepner Models Compared To Calabi-Yau Compactifications}, Ann. Phys. {\bf
  204} (1990)
1--51

\bibitem{Greene:1990ud}
B.~R. Greene and M.~R. Plesser, \emph{Duality In Calabi-Yau Moduli Space},
  Nucl. Phys. {\bf B338} (1990)
15--37

\bibitem{Kreuzer:1993uy}
M.~Kreuzer and H.~Skarke, \emph{ADE string vacua with discrete torsion}, Phys.
  Lett. {\bf B318} (1993) 305--314,
\href{http://www.arXiv.org/abs/hep-th/9307145}{{\tt hep-th/9307145}}

\bibitem{Gepner:1987hi}
D.~Gepner, \emph{String Theory on Calabi-Yau manifolds: The Three Generations
  case},
\href{http://www.arXiv.org/abs/hep-th/9301089}{{\tt hep-th/9301089}}

\bibitem{Pradisi:1996yd}
G.~Pradisi, A.~Sagnotti  and Y.~S. Stanev, \emph{Completeness Conditions for
  Boundary Operators in 2D Conformal Field Theory}, Phys. Lett. {\bf B381}
  (1996) 97--104,
\href{http://www.arXiv.org/abs/hep-th/9603097}{{\tt hep-th/9603097}}

\bibitem{Antoniadis:1999xk}
I.~Antoniadis, E.~Dudas  and A.~Sagnotti, \emph{Brane supersymmetry breaking},
  Phys. Lett. {\bf B464} (1999) 38--45,
\href{http://www.arXiv.org/abs/hep-th/9908023}{{\tt hep-th/9908023}}

\bibitem{FSS}
J.~Fuchs, A.~N. Schellekens  and C.~Schweigert, \emph{A matrix S for all simple
  current extensions}, Nucl. Phys. {\bf B473} (1996) 323--366,
\href{http://www.arXiv.org/abs/hep-th/9601078}{{\tt hep-th/9601078}}

\bibitem{Schellekens:1999yg}
A.~N. Schellekens, \emph{Fixed point resolution in extended WZW-models}, Nucl.
  Phys. {\bf B558} (1999) 484--502,
\href{http://www.arXiv.org/abs/math.qa/9905153}{{\tt math.qa/9905153}}

\bibitem{AnS}
C.~Angelantonj and A.~Sagnotti, \emph{Open strings}, Phys. Rept. {\bf 371}
  (2002) 1--150,
\href{http://www.arXiv.org/abs/hep-th/0204089}{{\tt hep-th/0204089}}

\bibitem{thesis}
L.~R. Huiszoon,
\emph{D-branes and O-planes in string theory: An algebraic approach {\rm PhD
  thesis, available on request}},

\bibitem{Fuchs:2004dz}
J.~Fuchs, I.~Runkel  and C.~Schweigert, \emph{TFT construction of RCFT
  correlators III: Simple currents},
\href{http://www.arXiv.org/abs/hep-th/0403157}{{\tt hep-th/0403157}}

\bibitem{BiM}
M.~Bianchi and J.~F. Morales, \emph{Anomalies and tadpoles}, JHEP {\bf 03}
  (2000) 030,
\href{http://www.arXiv.org/abs/hep-th/0002149}{{\tt hep-th/0002149}}

\bibitem{comments}
C.~Angelantonj, M.~Bianchi, G.~Pradisi, A.~Sagnotti  and Y.~S. Stanev,
  \emph{Comments on Gepner models and type I vacua in string theory}, Phys.
  Lett. {\bf B387} (1996) 743--749,
\href{http://www.arXiv.org/abs/hep-th/9607229}{{\tt hep-th/9607229}}

\end{thebibliography}\endgroup
\bibliographystyle{lennaert}

\end{document}